\documentclass[twocolumn,prl]{revtex4-1}
\usepackage{float}
\usepackage{times}
\usepackage{graphicx}
\usepackage{url}
\usepackage{amsmath}
\usepackage{amsfonts}
\usepackage{amssymb}
\usepackage[mathscr]{euscript}
\usepackage{cancel}
\usepackage{color}
\usepackage{dcolumn}
\usepackage{upgreek}
\usepackage{dsfont}

\newcommand\bra[1]{\langle #1 |}
\newcommand\ket[1]{| #1 \rangle}
\newcommand{\tlambda}{\tilde{\lambda}}
\begin{document}
\title{Fock-state stabilization and emission in superconducting circuits using dc-biased Josephson junctions}
\author{J.-R. Souquet, A.\,A. Clerk}
\affiliation{Department of Physics, McGill University, 3600 rue University, Montreal, Quebec, H3A 2T8, Canada }
\date{\today}
\begin{abstract}
We present and analyze a reservoir-engineering approach to stabilizing Fock states in a superconducting microwave cavity which does not require any microwave-frequency control drives.  Instead, our system makes use of a Josephson junction biased by a dc voltage which is coupled both to a principle storage cavity and a second auxiliary cavity.  Our analysis shows that Fock states can be stabilized with an extremely high fidelity.  We also show how the same system can be used to prepare on-demand propagating Fock states, again without the use of microwave pulses.  
\end{abstract}
\maketitle

\textit{Introduction--  }The ability to prepare, stabilize and ultimately transmit non-trivial quantum states is crucial to quantum information processing \cite{ladd2010quantum}.  In the presence of dissipation, state stabilization can be accomplished either via measurement-plus-feedback schemes (see e.g. \cite{calderbank1996good,nelson2000experimental,smith2002capture,cook2007optical,gillett2010experimental,sayrin2011real,vijay2012stabilizing,riste2013deterministic}), or autonomously, using quantum reservoir engineering (QRE) techniques \cite{poyatos1996quantum,Murch2012}.
There has been considerable progress in implementing QRE ideas in superconducting circuits, including experiments which have stabilized qubit states
\cite{Murch2012,shankar2013autonomously,Irfan2015} as well as photonic Fock states \cite{holland2015single} inside microwave frequency cavities.  These schemes are typically complex, requiring the use of several high-frequency microwave control tones.

In this work, we analyze an alternative approach to the stabilization of Fock states in a superconducting resonant cavity which requires no microwave control tones.
The starting point for our scheme is a setup discussed in a number of recent theoretical studies \cite{PhysRevLett.111.247001,PhysRevLett.111.247002,PhysRevB.92.054508} and experiments \cite{hofheinz2011bright,PhysRevB.90.020506,Altimiras,samkharadze2015high}, where a microwave cavity is coupled to a dc-biased Josephson junction.
The driven junction does not act like a qubit, but rather as a highly non-linear driving element.
That such a setup can produce nonclassical photonic states was first pointed out in Ref.~\cite{PhysRevLett.111.247002}, which showed that states could be produced having suppressed number fluctuations and a vanishing $g^{(2)}$ intensity correlation function.
While these states violate a classical Cauchy-Schwarz inequality, they are mixtures of the vacuum and the one photon Fock state, and have a limited fidelity with a pure Fock state.
As a result, they do not exhibit any negativity in their Wigner functions \cite{hakiogammalu2013quantum}.

\begin{figure}
	\centering
	\includegraphics[width=.45\textwidth]{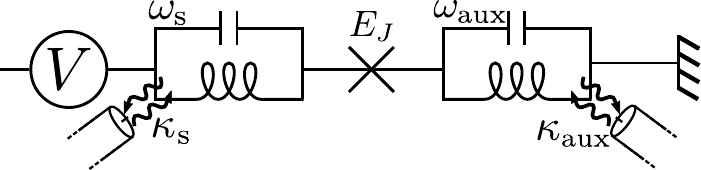}
	\caption{\label{fig:schematics} Schematic of the proposed set up.  A dc biased Josephson junction of energy $E_{J}$ is in series with two cavities.  Each cavity has an impedance close to $R_K = h / e^2$, and is damped via a coupling to a transmission line.}
\end{figure}

Here, we show how optimally coupling the junction to a second cavity (as depicted in Fig.~\ref{fig:schematics}) lets one transcend the limitations of the single cavity system, and prepare single-photon Fock states with an extremely high fidelity.
Such two-cavity-plus-junction systems have recently been realized experimentally \cite{hofheinz2011bright}, and have also been discussed theoretically
\cite{Fogelstrom2013,armour2015josephson,trif2015photon}, largely in the context of generating cavity-cavity correlations.
Armour et \textit{al.} \cite{armour2015josephson} found numerically that such a system could generate cavity states with weakly negative Wigner functions; they did not however discuss Fock state stabilization, or the particular mechanism we elucidate and optimize.

In addition to efficient and high-fidelity Fock state stabilization, we show that our setup can also play another crucial role:  it can act as an efficient means for producing propagating Fock states on demand.
In circuit QED setups, propagating single-photon states are usually generated by driving microwave cavities coupled to a qubit with high-frequency control pulses, see e.g.~\cite{houck2007generating}.
Among the many challenges in the standard approach is the requirement that the generated photon should be far detuned from the frequency of the control pulse \cite{eichler2011experimental,pechal2014s,kindel2015generation}.
Our system is capable of on-demand Fock state generation without any microwave control pulses:  one simply needs to pulse a dc control voltage.

\textit{Biased junction as a nonlinear drive--  }To set the stage for our two-cavity system, we start by quickly reviewing the physics of the
one-cavity version, as studied in  \cite{PhysRevLett.111.247002,PhysRevLett.111.247001}.
The system consists of a Josephson junction coupled to a cavity (frequency $\omega_{\rm c}$, modeled as an LC resonator), such that the voltage across the junction is the sum of an applied external dc voltage $V_{\rm dc}$ and fluctuating voltage $\hat V_{\rm cav}$ associated with the cavity mode.
The cavity is also coupled to a transmission line, which is treated (as standard) as a Markovian reservoir, and which gives rise to an energy damping rate $\kappa$.
We focus exclusively on temperature and voltages small enough that superconducting quasiparticles are never excited.

Working in an interaction picture at the cavity frequency, the Hamiltonian of the cavity plus biased junction is \cite{trif2015photon,meister2015resonators,PhysRevLett.111.247002,PhysRevLett.111.247001}
\begin{equation}\label{eq:hj}
	\hat H_{ J}= -\frac{E_{ J}}{2}\left(
	  e^{2i eV_{\rm dc}t}\hat D[\alpha (t)]+e^{-2i eV_{\rm dc}t}\hat D^\dagger[\alpha (t)]
	  \right),
\end{equation}
where $E_{J }$ is the Josephson energy, and $\hat D[ \alpha (t)]$ is the cavity displacement operator (corresponding to a time-dependent displacement
$\alpha(t)$).
It is defined in terms of the photon annihilation operator $\hat{a}$ as
\begin{equation}\label{eq:displacement}
	\hat D[ \alpha (t)] = e^{\alpha (t)\hat a^{\dagger}-\alpha^*(t)\hat a},\,\,
	\alpha (t)=	2\lambda e^{i \omega_{\rm c} t}.
\end{equation}
$\lambda$ determines the amplitude of the zero-point voltage fluctuations in the cavity, and is given by $\lambda=\sqrt{\pi e^2 Z/h }$ where $Z$ the impedance of the LC resonator.

From the point of view of the cavity, $\hat{H}_J$ describes a highly-nonlinear (but coherent) cavity drive \cite{PhysRevLett.111.247002}: each term tunnels a Cooper-pair, and also displaces the cavity state by $\pm \alpha(t)$.
To see clearly how these displacements can result in Fock state generation, we follow a different route from \cite{PhysRevLett.111.247002}, and
express $\hat D[\alpha(t)]$ directly in the Fock basis (see, e.g.~\cite{cahill1969ordered}):
\begin{multline}\label{eq:DT}
	\hat D[\alpha(t)] = \sum_{n=0}^{\infty}\sum_{l=0}^{\infty}  w_{n,n+l}[\lambda]\, \ket {n} \bra{n+l}  e^{ - i l \omega_{\rm c} t }\\
		+\sum_{l=1}^{\infty} (-1)^l\ket {n+l} \bra{n}   w_{n+l,n}[\lambda]e^{ i l \omega_{\rm c} t }.
\end{multline}
Here, the transition amplitude $w_{n,n+l}[\lambda]$ is nothing more than a generalized Frank-Condon factor.
For $l \geq 0$ we have:
\begin{equation}
	w_{n+l,n}[\lambda]
	=  e^{-2\lambda^2 } (2\lambda)^{l} \sqrt{\tfrac{ n!}{ (n+l)!}} L_{n}^{(l)}(4\lambda^2),
\end{equation}
where $L_{n}^{(k)}$ is a Laguerre polynomial \footnote{By convention, $1/(l+n)!=0$ for $l+n<0$}, and $w_{n+l,n}[\lambda]=w_{n,n+l}[-\lambda]$.

\begin{figure}
	\centering
\includegraphics[width=.45\textwidth]{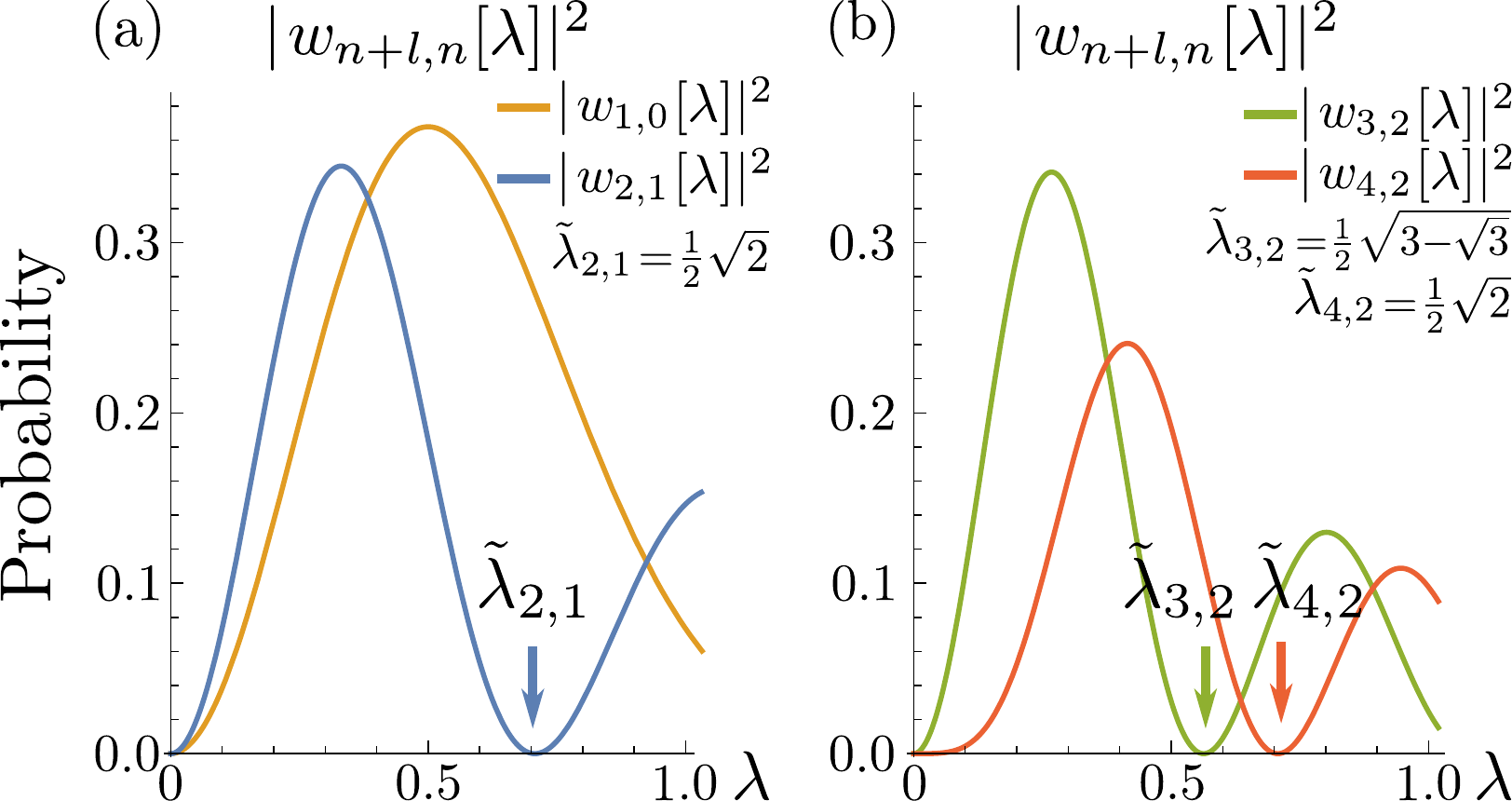}
	\caption{\label{fig:qnl}Matrix elements $w_{n+l,n}[\lambda]$ for junction-driven cavity Fock state transitions, as a function of the zero-point voltage fluctuation amplitude $\lambda$ for different values of $n$ and $l$.  These functions are highly non-linear with respect to $\lambda$, and can cancel for specific values.  Roots of $w_{n+l,n}[\lambda]$, denoted $\tlambda_{n+l,n}$, are indicated in the figure.  }
\end{figure}

As is well known, Frank-Condon factors are highly nonlinear functions of the magnitude of the displacement (here set by $\lambda$), and can even exhibit zeros \cite{smith1968cancellation};  the behavior of relevant factors is shown Fig.~\ref{fig:qnl}.
We let $\tlambda_{n+l,n}$ denote the smallest value of $\lambda$ which makes $w_{n+l,n}[\lambda]$ vanish:
\begin{equation}
	w_{n+l,n}[\tlambda_{n+l,n}] = w_{n,n+l}[\tlambda_{n+l,n}] = 0.
\end{equation}

The route to preparing single Fock states now seems clear:  by tuning the value of both $V_{\rm dc}$ and  $\lambda$ (via the cavity impedance $Z$), one can arrange for the effective driving of the cavity by the the junction to shut off when a given Fock state is reached \cite{holland2015single}.
For concreteness, suppose we chose $V_{\rm dc} = k \Omega / 2e$ ($k$ an integer), so that to leading order, Cooper-pairs can only tunnel by emitting or absorbing $k$ cavity photons.
If we take $\omega_c\gg E_J, \kappa$ we can restrict attention to these processes and make a rotating-wave approximation (RWA) to our full Hamiltonian.  Within the RWA we have:
\begin{equation}\label{eq:hjrwa}
	 \hat H_{\rm RWA} =(-1)^{k+1} \frac{E_J}{2} \sum_{n=0}^{\infty} w_{n+k,n}[\lambda] \ket{n+k}\bra n+{\rm h.c.}.
\end{equation}
Consider the simplest case where $k=1$, and each resonantly-tunnelling Cooper pair emits or absorbs a singe cavity photon.
If we then set $\lambda = 1 / \sqrt{2} \equiv \tlambda_{2,1}$, there is no matrix element in Eq.~(\ref{eq:hjrwa}) for a transition from $1$ to $2$ cavity photons.
As first discussed in Ref.~\cite{PhysRevLett.111.247002}, the junction induced drive now can add at most one photon to the cavity, implying that the system effectively acts like a driven two-level system.

The cavity steady-state is found by solving the Linblad master equation for the density matrix of the cavity $\hat \rho_{\rm c}$,
\begin{equation}\label{eq:meq1}
	\dot{\hat \rho}_{\rm c} = -i [\hat H_{\rm RWA},\hat \rho_{\rm c}] + \kappa {\cal L} [\hat a]\hat \rho_{\rm c},
\end{equation}
which includes the dissipation from the (zero-temperature) transmission line; here, ${\cal L}[\hat a]\hat\rho=\hat a \hat\rho \hat a ^{\dagger}-\tfrac12\{\hat \rho\hat n+\hat n\hat \rho\}$.
When $\lambda$ is set to $\tlambda_{2,1}$, the stationary intracavity state can be termed non-classical, in that it results in a vanishing $g^{(2)}$ intensity-intensity correlation function \cite{PhysRevLett.111.247002}.
This simply reflects the fact that there is zero probability for having two or more photons in the cavity.
We are still far however from our goal of producing a single-photon Fock state.
As the cavity is effectively a driven two-level system, population inversion is impossible, and at best the steady state is an incoherent mixture having an equal probability of vacuum and single photon.
We stress that such a state exhibits no negativity in its Wigner function.

\textit{Fock state stabilization--  }
Heuristically, the poor performance of the single-cavity setup is easy to understand:  even if we eliminate the matrix element for $\ket 1 \rightarrow \ket 2$ transitions, the junction-driven cavity continues to oscillates back and forth between the vacuum and the $\ket 1$ Fock state (eventually relaxing into a mixed state).
To achieve true Fock state generation, one needs to shut off the oscillation dynamics when the system is in the $\ket 1$ state.
As we now discuss, this can be achieved rather simply by coupling the junction to a second ``auxiliary" cavity (see Fig.~\ref{fig:schematics}) whose damping rate $\kappa_{\rm aux}$ is taken to be sufficiently large.

In what follows, we denote quantities for the main ``storage" cavity with a subscript ``s", while auxiliary cavity quantities have a subscript ``aux";  Fock states of the two-mode system are denoted $\ket{n,m}=\ket n_{\rm s}\otimes \ket m_{\rm aux}$.
Working in an interaction picture with respect to the free cavity Hamitonians, the system Hamiltonian is ( here and throughout the text, $\hbar=1$),
\begin{equation}
	\hat {\cal H}_{ J}=-\tfrac{E_{ J}}2 e^{2i e V_{\rm dc}t} \hat {\cal D}[\alpha_{\rm s}(t),\alpha_{\rm aux}(t)]+{\rm h.c.}.
\end{equation}
Here $\hat{\cal D}[\alpha_{\rm s}(t),\alpha_{\rm aux}(t)]=\hat D_{\rm s}[\alpha_{\rm s}(t)]\otimes\hat D_{\rm aux}[\alpha_{\rm aux}(t)]$ is the tensor product of displacement operators for each cavity, with respective displacement amplitudes $\alpha_{j}(t)=2\lambda_{j} e^{i \omega_{j} t}$.

\begin{figure}[t]
	\centering
\includegraphics[width=.47\textwidth]{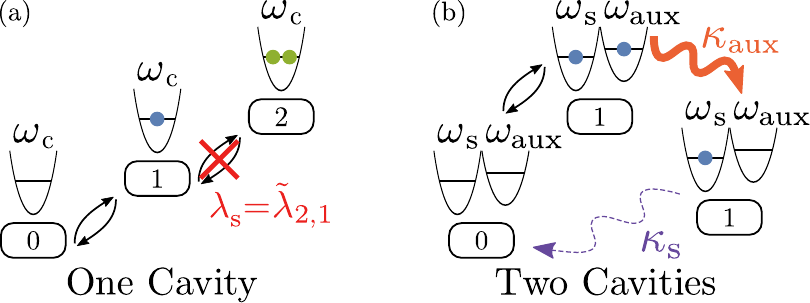}
	\caption{\label{fig:PP_v2} Schematic depictions of biased-junction cavity pumping processes.  The boxed digit indicates the number of Cooper pairs that have tunnelled, whereas dots in the parabolas indicate intracavity photon number.  a) Single cavity setup for an impedance yielding $\lambda = \tilde{\lambda}_{2,1}$.  Cooper-pair tunnelling can take the cavity between the 0 and 1 photon states, but transitions to the 2-photon state are blocked.
	 b) Two-cavity setup for Fock state stabilization, where $\lambda_{\rm s} = \tlambda_{2,1}$.  Starting from vacuum, Cooper-pair tunneling can cause oscillations between the $\ket{0,0}$ and $\ket{1,1}$ photon Fock states.  Photon decay from the aux cavity however freezes the system into the desired $\ket{1,0}$ state.  When the storage cavity photon decays (due to $\kappa_{\rm s}$), the cycle repeats.  }
\end{figure}

For the auxiliary cavity to play the desired role, we tune the voltage so that $V_{\rm dc}=(\omega_{\rm s}+\omega_{\rm aux})/2e$.  Cooper-pair tunnelling thus requires simultaneously emitting a single photon to each cavity (or absorbing a photon from each cavity).
We also tune the storage cavity impedance so that $\lambda_{\rm s} = \tlambda_{2,1}$; we stress that no special tuning of $\lambda_{\rm aux}$ is needed.
The resulting dynamics is sketched in Fig.~\ref{fig:PP_v2}b.
Similar to the single-cavity system, the effective driving from the biased junction couples $\ket{0,0}$ to $\ket{1,1}$, but
not to states with higher photon number.
We would seem yet again to have an effective two-level system, and might expect coherent oscillations between these two states.
However, the large damping rate $\kappa_{\rm aux}$ of the auxiliary cavity prevents this:  if the system is in the $\ket{1,1}$ state, $\kappa_{\rm aux}$ will cause a rapid decay to $\ket{1,0}$.
In the absence of storage cavity damping, the system is then effectively stuck:  Cooper-pair tunnelling against the voltage is impossible (as there are no photons in the aux cavity), while tunnelling with the voltage is impossible as there is no matrix element connecting $\ket{1,0}$ and $\ket{2,1}$.
Including storage cavity-damping does not ruin the physics:  if the storage cavity photon leaks out, one is back in the vacuum state, and the process starts again.
One thus sees the possibility for having a steady state that has a high fidelity with the state $\ket{1,0}$, i.e. a stabilized single-photon state in the storage cavity.

\begin{figure}
		\centering
		\includegraphics[width=.482\textwidth]{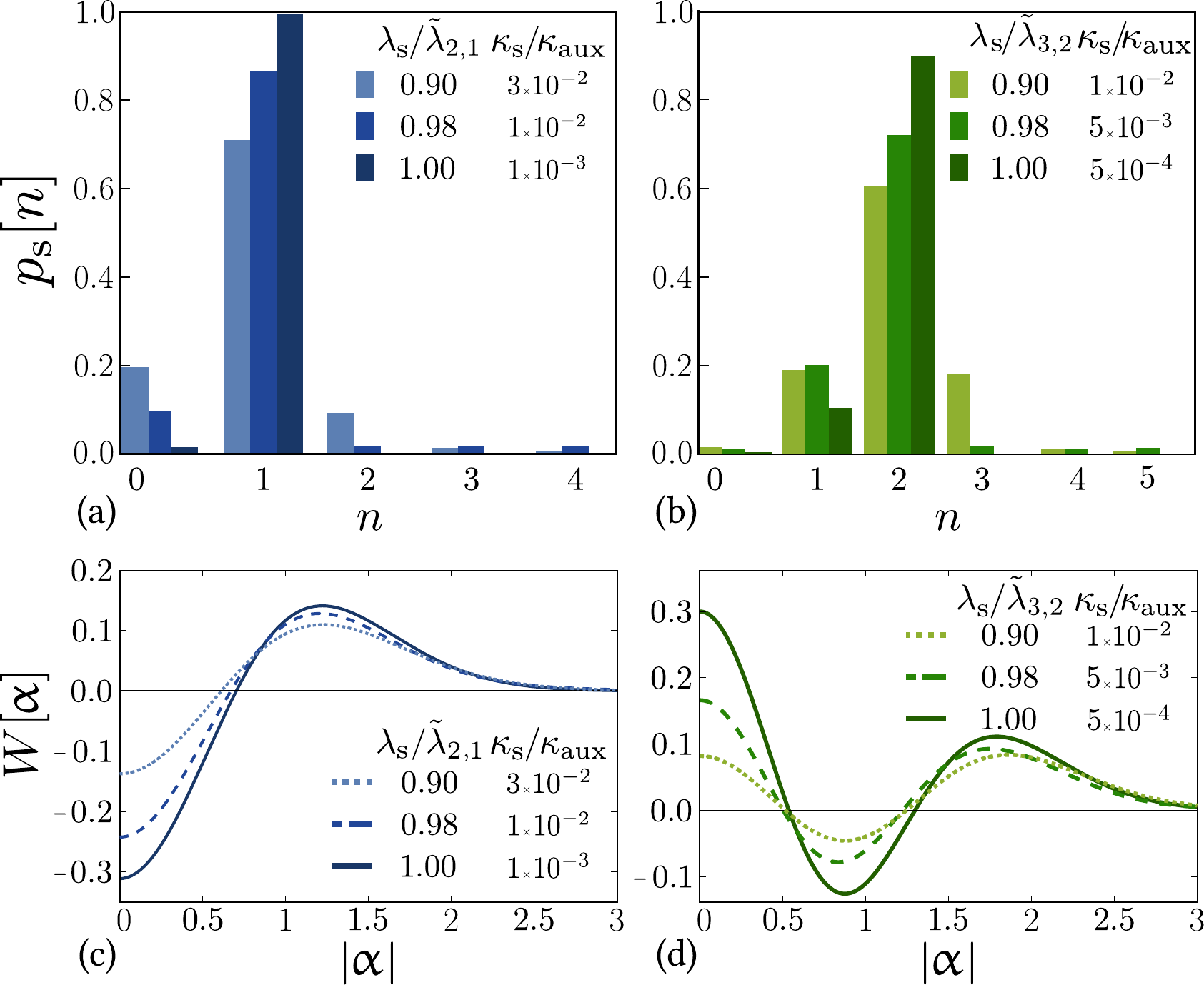}
	\caption{\label{fig:2-4Photons} Steady-state probability $p_{\rm s}[n]$ for the storage cavity to be in a Fock state $\ket{n}$, as obtained
	from the RWA master equation in Eq.~(\ref{eq:RWAMaster}).  We take $E_{ J}=\kappa_{\rm aux} \ll \omega_{\rm s}, \omega_{\rm aux}$,
	and $\lambda_{\rm aux} = 1/2$; $\lambda_{\rm s}$ and $\kappa_{\rm s}$ are indicated in the plots.
	(a) Probabilities when $\lambda_{\rm s}$ is tuned close to $\tlambda_{2,1}$ (for stabilizing $\ket{1}$).  (b)  Same, for
		$\lambda_{\rm s}$ tuned close to $\tlambda_{3,2}$ (for stabilizing $\ket{2}$).  Even with imperfect tuning, the target Fock state is stabilized with a high fidelity.
		(c) and (d) Corresponding Wigner function $W[\upalpha]$ for the storage cavity state, showing that negative values are obtained; note that $W[\upalpha]$ is rotationally invariant in all cases.
		}
\end{figure}

To make the above picture quantitative, we focus on a voltage $V_{\rm dc}=(\omega_{\rm aux}+\omega_{\rm s})/2e$, and make a RWA on our Hamiltonian, yielding:
\begin{multline}\label{eq:h2jrwa}
	\hat {\cal H}_{\rm RWA} =\frac{E_{ J}}{2}
		\left(  \sum_{n_{\rm s}=0}^\infty   w_{ n_{\rm s}+1, n_{\rm s}}[\lambda]
			 \ket{n_{\rm s}+1} \bra{n_{\rm s}}  \right)\\
\otimes\left(  \sum_{n_{\rm aux}=0}^\infty   w_{ n_{\rm aux}+1, n_{\rm aux}}[\lambda] \ket{n_{\rm aux}+1} \bra{n_{\rm aux}} \right)+ {\rm h.c.}.
\end{multline}
Each cavity is also coupled to its own zero temperature bath, and the reduced density matrix $\hat \rho$ of the two cavities obeys the Linblad master equation:
\begin{equation}
	\dot{\hat \rho} = -i [\hat {H}_{\rm RWA}, \hat \rho] +\kappa_{\rm aux}{\cal L}[\hat a_{\rm aux}] \hat \rho
	+\kappa_{\rm s}{\cal L}[\hat a_{\rm s}] \hat \rho.
	\label{eq:RWAMaster}
\end{equation}

Consider first the ideal case, where $\lambda_{\rm s}$ is tuned perfectly to equal $\tlambda_{2,1}$.  In the relevant limit $\kappa_{\rm s} \ll \kappa_{\rm aux}$, the steady state probability that the system is in the desired state $\ket{1,0}$ is
\begin{equation}\label{eq:gamma}
	\bra{1,0} \hat{\rho} \ket{1,0} \simeq \frac{\Gamma }
	{\Gamma+ \kappa_{\rm s}},
	\hspace{0.3 cm}
	\Gamma = \frac{ \left(
		E_{J} w_{1,0}[\lambda_{\rm s}] w_{1,0}[\lambda_{\rm aux}] \right)^2 }
		{\kappa_{\rm aux}},
\end{equation}
where $\Gamma$ plays the role of an effective pumping rate from $\ket{0,0}$ to $\ket{1,0}$, and we have dropped terms as small as $\kappa_{\rm s} / \kappa_{\rm aux}$.  The probability to be in the desired state tends to $1$ in the limit $\Gamma \gg \kappa_{\rm aux} \gg \kappa_s$.  The large $\ket{1,0}$ population here is analogous to the population inversion possible in a driven three-level system \cite{walls2007quantum}.

The above process can be efficient even if the storage cavity impedance is not perfectly tuned.  Suppose $\lambda_{\rm s}=\tlambda_{2,1}+\varepsilon$ with $\varepsilon\ll1$.  Assuming again $\kappa_{\rm s}\ll \kappa_{\rm aux}$, one finds that the probability to be in $\ket{1,0}$ is modified to be:
\begin{equation}\label{eq:theta}
	\bra{1,0} \hat{\rho} \ket{1,0}  \simeq
	 \frac{\Gamma}{\Gamma + \kappa_{\rm s} + 4\varepsilon^2 \Gamma^2 / \kappa_{\rm s} }.
\end{equation}
Because of the imperfect tuning of $\lambda_{\rm s}$, it is no longer advantageous to have $\Gamma \gg \kappa_{\rm s}$ (i.e.~large $E_{J}$), as transitions to higher states will corrupt the dynamics.  Eq.~\eqref{eq:theta} suggests that an optimal choice would be to have $\kappa_{\rm s} =2\varepsilon \Gamma$ (while still maintaining $\kappa_{\rm aux} \gg \kappa_{\rm s}$).

To complement the above analytical results, we have performed a full numerical simulation of the RWA master equation in Eq.~(\ref{eq:RWAMaster}) using the QuTiP package \cite{Johansson20131234}.  Fig.~\ref{fig:2-4Photons}a shows the stationary storage cavity Fock state distribution as a function of $\lambda_{\rm s}$, and demonstrates that high-fidelity Fock state generation is possible despite an imperfect tuning of the storage cavity impedance.  The fidelity is sufficient to give rise to storage-cavity Wigner function that exhibit large amounts of negativity (as shown in Fig.~\ref{fig:2-4Photons}c).

The above protocol can also be used to stabilize higher Fock states with a good fidelity.  One keeps the voltage set to $V_{\rm dc} = (\omega_{\rm s} + \omega_{\rm aux})/2e$, but now tunes the storage cavity impedance such that $\lambda_{\rm s} = \tlambda_{n+1,n}$ for some chosen $n > 1$.  The system dynamics will now effectively get stuck in the state $\ket{n,0}$.  Numerical results for the case $n=2$ are shown in Fig.~\ref{fig:2-4Photons}b and d.

We stress that the general scheme here can be viewed as an example of reservoir engineering \cite{poyatos1996quantum}, with the biased junction and auxiliary cavity acting as an effective dissipative environment which stabilizes the storage cavity in the desired Fock state.  In that respect, our protocol has similarities to the Fock-state stabilization scheme described and implemented in Ref.~\cite{holland2015single}.  In that work, the engineered reservoir also involved an auxiliary cavity, but used a qubit and two microwave control tones (or more, if the target Fock state has $n>1$).  In our work, the auxiliary cavity is still there, but the microwave control tones and qubit have been replaced with a Josephson junction biased by a dc-voltage.  Note that conversely to Ref.~\cite{holland2015single}, reaching higher Fock states does not require additional resources.

\begin{figure}
\includegraphics[width=.47\textwidth]{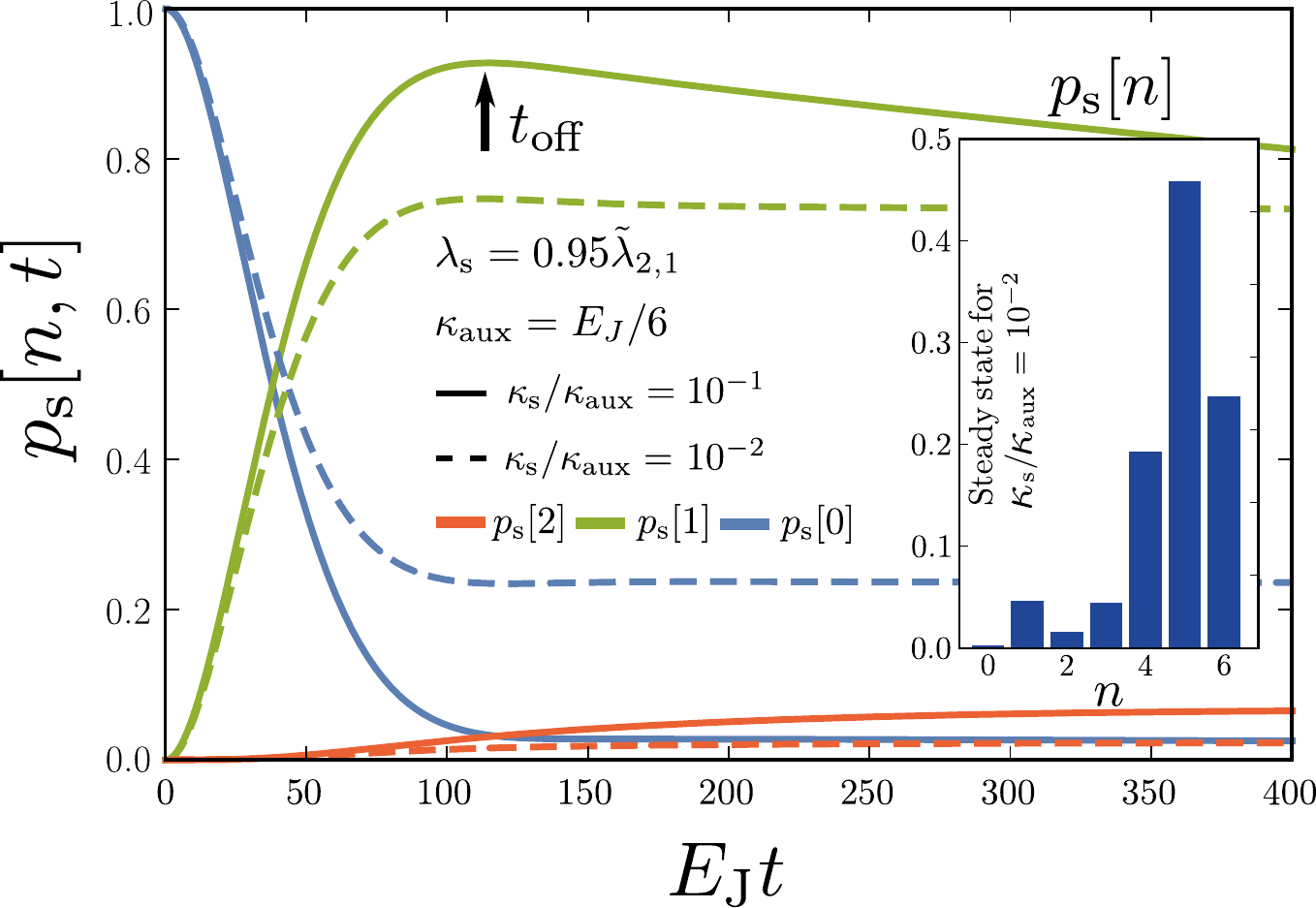}
	\caption{\label{fig:MC} Time dependence of storage cavity photon number occupancies $p_s[n,t]$, where the bias $V_{\rm dc} =(\omega_{\rm aux}+\omega_{\rm s})/2e$ is turned on at $t=0$, and the cavities start from vacuum.  Dashed line: $\kappa_{\rm s}$ chosen to optimize the steady-state value of $p_{\rm s}[1,t]$ (the Fock-state stabilization protocol).  Solid line: alternate choice of $\kappa_{\rm s}$ which optimize $p_{\rm s}[1,t]$ at intermediate times (pulsed protocol).  In this latter protocol, the voltage can be turned of near $t_{\rm off}$, resulting with high-probability in the production of a propagating Fock state in the transmission line coupled to the storage cavity.  In both cases, we assume the realistic situation where the cavity impedance has not been tuned perfectly (here, $\lambda = 0.95 \tilde \lambda_{21}$ ).  The inset shows $p_{\rm s}[n,t \to \infty]$ for the pulsed protocol. }
\end{figure}

\textit{Itinerant Fock states on demand-- }
The above protocol for stabilizing Fock states can naturally be used to produce propagating Fock states:  after the desired Fock state has been stabilized,
one simply turns off the dc bias voltage when the itinerant photon is desired, and the storage cavity Fock state will be emitted into the transmission line coupled to it (in a temporal mode having an exponential profile \cite{yin2013catch,maxwell2012storage}).  An alternate strategy is to exploit the transient dynamics of our scheme, and produce an itinerant Fock state with a pulsed dc voltage.  In this case, one optimizes parameters to have a high-fidelity intra-cavity Fock state at an intermediate time, as opposed to in the long-time steady state.  The voltage is then turned off at this intermediate time.  This makes it possible to achieve a high fidelity Fock state even if the tuning of the storage cavity impedance is not perfect, i.e.~$\lambda_{\rm s} \neq \tlambda_{2,1}$.  In addition, by choosing parameters such that the oscillation dynamics associated with Cooper-pair tunnelling is slightly overdamped, one can have a protocol which is relatively insensitive to the timing of the voltage pulse.

In Fig.~\ref{fig:MC}, we show a comparison between the steady-state stabilization protocol and this pulsed protocol, where $\lambda_{\rm s}$ deviates by $5 \%$ from the ideal value $\tlambda_{2,1}$ needed for single-photon generation.  By using a smaller value of $\kappa_{\rm s}$ than the choice that optimizes the steady-state single-photon probability, one obtains a much higher fidelity with a single photon Fock state at intermediate times.  Further, the maximum probability for having a single storage-cavity photon is a rather broad function of time, meaning that one does not need precise control of the shut-off time of the dc-voltage.  This is in stark contrast to standard protocols for preparing a Fock state using two-level system dynamics (e.g.~in the one-cavity version of our system), where one needs a precise control of the duration of the pulses control.

\textit{Conclusion-- }
We have shown how a system where a voltage-biased Josephson junction is coupled to two cavities can be used to stabilize Fock states with a high efficiency.  While our approach does require one to carefully tune the impedance of the main storage cavity, it does not require any microwave-frequency control tones.  We also discussed how the same setup could be used to produce itinerant single-photons on demand.

This work was supported by NSERC.

\bibliography{biblio}

\end{document}